\documentclass[a4paper]{eccomas_paper-2024}
\usepackage{graphicx}
\usepackage{hyperref}
\usepackage{graphicx}
\usepackage{amsmath}
\usepackage{amsfonts}
\usepackage{amssymb}
\usepackage{anyfontsize}

\usepackage{standalone}

\usepackage[numbers]{natbib}

\title{COMPARISON OF INTEGRATION METHODS FOR CUT ELEMENTS}

\author{G. H. TEIXEIRA$^{1}$, M. LOIBL$^{2}$, B. MARUSSIG$^{1}$}

\heading{G. H. TEIXEIRA, M. LOIBL, B. MARUSSIG}

\address{$^{1}$ Institute of Applied Mechanics\\
Graz University of Technology,\\
Technikerstraße 4/II, 8010, Graz, Austria\\
e-mail: teixeira@tugraz.at
\and
$^{2}$ Institute of Mechanics and Statics,\\
University of the Bundeswehr Munich.\\
Werner-Heisenberg-Weg 39, 85577, Neubiberg, Germany\\
e-mail: michael.loibl@unibw.de
}

\keywords{Integration, Cut Elements, IsoGeometric Analysis}

\abstract{}

\begin{document}
\thispagestyle{empty}
Using an interface inserted in a background mesh is an alternative way of constructing a complex geometrical shape with a relative low meshing efforts. However, this process may require special treatment of elements cut by the interface.  
Our study focuses on comparing the integration of cut elements defined by implicit and parametric curves. 
We investigate the efficiency and robustness of open-source tools such as Algoim \cite{Saye2022}(a library for quadrature on implicitly defined geometries) and Ginkgo \cite{Antolin2019}(a library for isogeometric analysis on Boolean operations with a parametric description) with numerical examples computing the area defined by the interface and benchmarks for 2D elasticity problem using the open-source code GeoPDEs \cite{Vazquez2016}.
It is concluded that none of the two interface descriptions is preferable with respect to the quality of the integration. Thus, the choice of the interface type depends only on the studied problem and the available curve description, but not on the numerical aspects of the integration.

\section{INTRODUCTION}
\begin{figure}[t]
        \centering
        \includegraphics[scale=1.2]{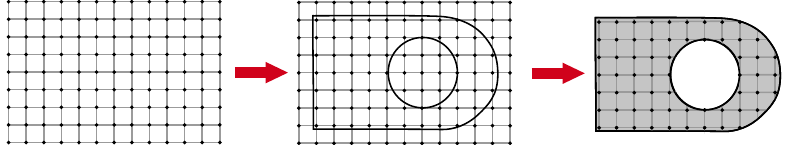}
        \caption{Scheme of domain definition with an interface}
        \label{fig1}
    \end{figure}

    Modelling the geometry and discretising the domain of a problem with a mesh is an important step in the process of solving any boundary value problem. Regarding the construction of a geometric representation, a common way to do it is to define a parametric shape and map this geometry into the physical domain. However, finding a suitable parametrization for complex geometries is in general not an easy task. One way to mitigate this problem is to define the geometry through a background mesh, where an interface specifies a region to be visible. This approach allows us to analyse any geometry by simply selecting a part of the background mesh to define the domain, as can be seen in the scheme presented in Figure \ref{fig1}. This concept is present in methods such as the Immersed Boundary Method \cite{Wei2021} or the Finite Cell Method \cite{Schillinger2015}.
    
    In the process of selecting the domain, some elements are cut by this interface. In consequence, among several challenges, the numerical integration of cut elements is an important point to be verified \cite{Marussig2018}. In particular, we would like to investigate, how the quadrature points are distributed over the cut elements, and we would like to observe how this is affected by the type of the interface used. Thereby, we compare parametric and implicit interface descriptions for cut elements. Depending on the type of interface an appropriate quadrature scheme is applied to define the points and weights in each cut element. In particular, in this work, for parametric curve interfaces, a reparametrisation of the element is performed and the quadrature is obtained as in \cite{Antolin2019}. Whereas for implicit curve interfaces, an approximation of the curve by a Bézier curve is carried out and the corresponding points and weights are defined as in \cite{Saye2022}. 
\section{INTERFACE REPRESENTATIONS}   
\subsection{Parametric curves}
A parametric curve is defined by the map from the parameter domain to the physical domain, where each point in the parameter domain has its own image in the physical domain. Bézier curves, B-splines, and NURBS are examples of parametric curves. In the comparisons presented here, NURBS curves $C(\xi)$ are used to create the interface, which can be defined by the linear combination of NURBS basis functions $R_{j,p}$ of degree $p$ scaled by control points $\mathbf{c}_i$
    \begin{equation}
         \mathbf{C}(\xi)=\sum_{j=1}^n R_{j,p}(\xi)\mathbf{c}_i.
    \end{equation}
    
At least one edge of a cut element is defined by this interface. And for such elements where the interface is defined by a NURBS curve or portion of it, we employ the open-source code Ginkgo \cite{Antolin2019} to compute the quadrature points and weights. In this approach, the first step to obtain the quadrature points is defined by the reparametrisation of the element cut by the interface. However, sometimes the reparametrisation of an element with only one B-spline surface is not possible. So, the cut element is divided into several tiles, where each of them can be reconstructed using B-spline surfaces. Once this is done, the quadrature points from a reference element are mapped to each one of these small parts of the reparametrised element.

\subsection{Implicit curves}
An implicit curve is defined by a locus of points for which an equation takes a specific value. The points with the same value form isolines and, for convenience,  the zero set is usually chosen as the isoline that implicitly describes the curve. These points are defined to fulfil the condition 
     \begin{equation}
         f(x,y)=0.
    \end{equation}
The use of implicit curves to define the domain can be applied to solve topology optimisation problems via the Level-set Method \cite{Allaire2002}. In order to compute the integral in implicitly defined domains, we apply the open-source code Algoim \cite{Saye2022} to determine quadrature points and weights. In this approach, the implicit curve is approximated by a Bézier curve and then the algorithm is used to obtain the quadrature rule for domains implicit defined by multivariate polynomials. The quadrature points and weights are obtained by dividing the domain in pieces where in each one of them the appropriate quadrature scheme is applied for one-dimensional integrals \cite{Saye2022}.

\section{METHODOLOGY}
\begin{figure}[t!]
        \centering
        \includegraphics[]{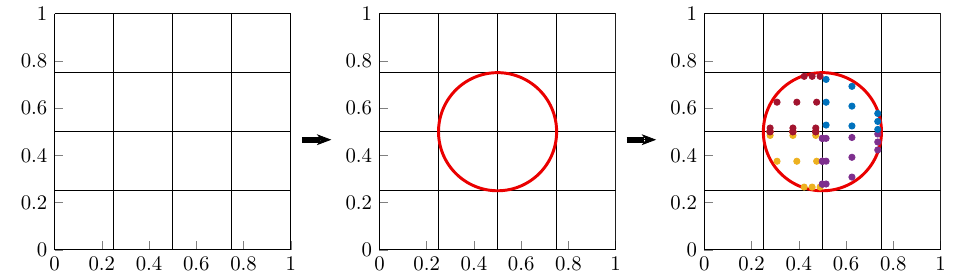}
        \caption{Procedure for obtaining the quadrature points}
        \label{fig2}
    \end{figure}

The process of defining the quadrature rule on the cut elements begins with the definition of a background mesh and is followed by the insertion of an interface that creates the domain of interest.  The difference between the approaches considered here is related to how these interfaces are defined for an identical geometry. Once the interface has been defined, the elements are identified as active and non-active. The non-active elements are neglected in the computation and the active elements are divided into two subgroups cut and uncut elements. For the uncut elements, a Gaussian quadrature is defined in a reference element and then mapped to the physical element, while for the cut elements the appropriate approach is applied depending on the type of interface. Figure \ref{fig2} describes the steps to obtain the quadrature rule. In a pre-existing background mesh, an interface defines the domain where, depending on the type of interface, the quadrature points are provided using the appropriate open-source code (Algoim \cite{Saye2022} or Ginkgo \cite{Antolin2019}), which in sequence allows the numerical integration to be performed by simply evaluating the function at the quadrature points and scaling by the corresponding quadrature weights.
In order to compare the different tools, first we consider the following verification settings:
    \begin{itemize}
        \item Computation of the area enclosed by an interface for different refinement levels; and
        \item computation of the area enclosed by an interface for a fixed refinement level while the geometry moves.
    \end{itemize}
With these two settings we can verify not only the accuracy but also the robustness of the two approaches comparing the relative error with respect to the size of mesh and the relative error with respect to a given step.
Once the behaviour of the tools is observed, we investigate how the definition of the interface could affect the solution of 2D elasticity problems, where a background mesh is defined using B-splines. For this, the open-source code for Isogeometric Analysis GeoPDEs \cite{Vazquez2016} is used. In this situation, two different scenarios of cut curves are set up, one interface with no boundary conditions applied to it and one interface with boundary conditions applied to it. 
    
\section{NUMERICAL RESULTS}
    \subsection{Area computation}
    \begin{figure}[t!]
        \centering
        \includegraphics[]{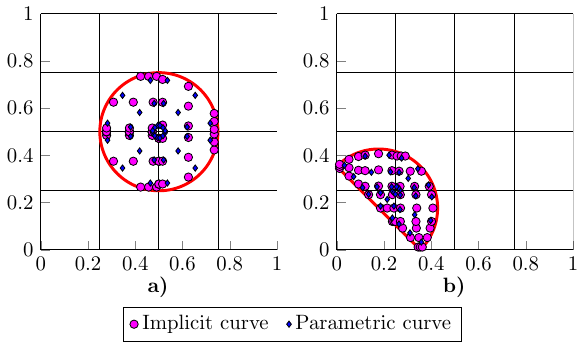}
        \caption{Quadrature points: a) circle example; b) semicircle example}
        \label{fig3}
    \end{figure}
    \begin{figure}[b!]
        \centering
        \includegraphics[]{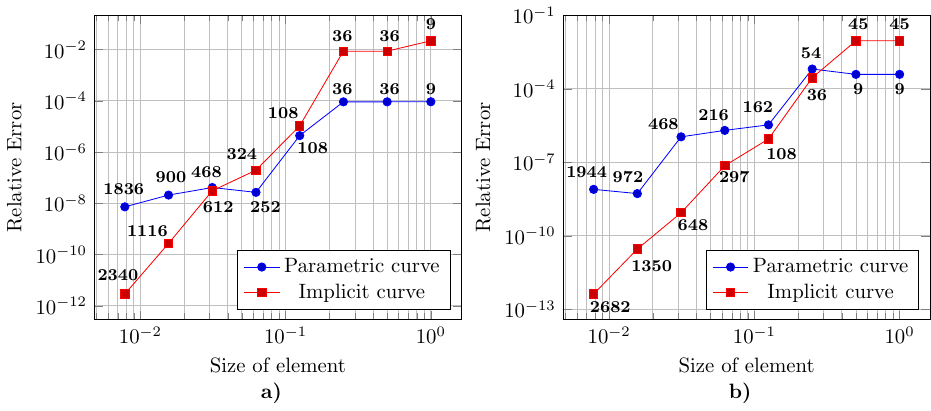}
        \caption{Convergence study: a) circle example; b) semicircle example}
        \label{fig4}
    \end{figure}
    Our first comparison study considers simply the computation of the area enclosed by an interface. In this test, we verify the accuracy by computing the relative error for a fixed interface while refining the background mesh. The first example is defined by a circle with radius $R=0.2$ and centered at $(0.5,0.5)$. In this example, the geometry is fixed using a single interface and the mesh is refined by h-refinement. Figure \ref{fig3}a shows the shape and quadrature points for a mesh of size $h=0.25$, which results in 36 quadrature points for both interface types. While the relative error with respect to the size of the elements is shown in Figure \ref{fig4}a.
    The second example is a semicircle, which is built using two interfaces. In particular, the interface using implicit curves is defined by the intersection of a straight line $y=-x+0.3536$ with a circle with radius $R=0.25$ and centered at (0.1768,0.1768). The mesh is refined by $h$-refinement and Figure \ref{fig3}b shows the quadrature points obtained for the same element size as the circle example. Figure \ref{fig4}b shows the relative error and the number of quadrature points needed for each size of mesh. 
    
    \subsection{Moving interfaces}
    \begin{figure}[t!]
        \centering
        \includegraphics[]{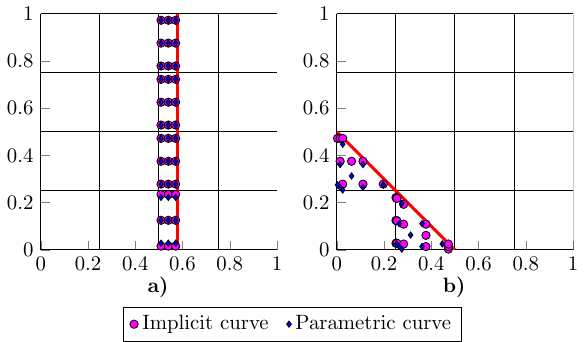}
        \caption{Quatrature points: a) Line ; b) Triangle}
        \label{fig5}
    \end{figure}
    \begin{figure}[b!]
        \centering
        \includegraphics[]{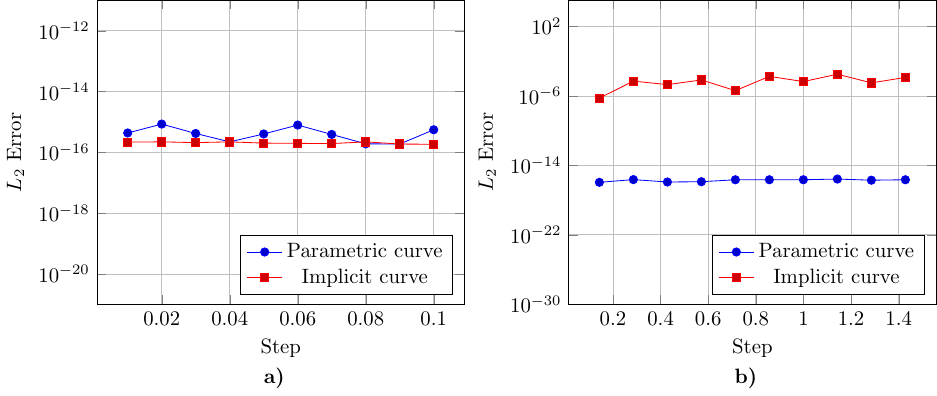}
        \caption{Robustness study: a) line; b) triangle}
        \label{fig6}
    \end{figure}
    Another aspect verified in this part of the work is the robustness of the approaches. For this, we look at different situations where it might be difficult to deal with the integration. This is done by setting a mesh of size $h=0.25$ and moving an interface around, trying to find positions or situations where the codes are crashing. We investigate two situations: The first one is defined by a straight vertical line moving along the horizontal direction, as shown in Figure \ref{fig5}a. While the second is given by a triangle in which the node $(0,0.5)$ rotates clockwise with respect to the origin, while the nodes at $(0,0)$ and $(0.5,0)$ remain fixed, as shown in Figure \ref{fig5}b.
    Figure \ref{fig6} shows the relative error with respect to the given steps for both examples. For the line example, the step represents the distance with respect to the first position of the line, which is at $x=0.5$, and for the triangle example, the step represent the angle between the line that connects the origin to the moving node and its original position, the line that connects the origin to the node $(0,0.5)$. As it is possible to see, none of the two approaches fails for any of the investigated steps.
    \begin{figure}[t!]
        \centering
        \includegraphics[]{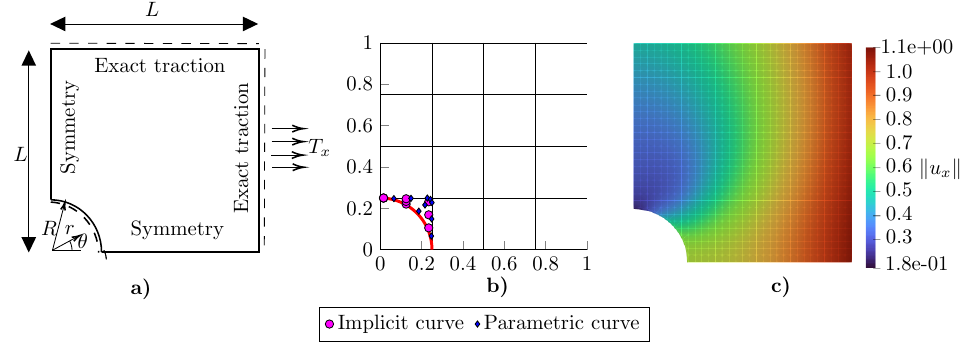}
        \caption{Infinite plate with a hole: a) problem scheme; b) quadrature points; c) displacement magnitude}
        \label{fig7}
    \end{figure}
    
    \subsection{Elasticity benchmarks}
    The 2D elasticity problems are solved using the open-source code for Isogeometric Analysis GeoPDEs \cite{Vazquez2016} with quadrature points and weights provided by Algoim \cite{Saye2022} and Ginkgo \cite{Antolin2019} .The geometries are defined by the intersection between a B-spline background mesh and one of the two types of an interface. And it is considered $E=1$ and $\nu=0.3$.

    \subsubsection{Infinite plate with a hole}
    The first example is the well-known benchmark of an infinite plate with a hole subjected to a traction $T_x=10$ at infinity, which can be modeled by a quarter of the geometry due to symmetry. To do this, the background mesh $\Omega=[0,1]\times[0,1]$ is cut by a circle with radius $R=0.25$, as shown in Figure \ref{fig7}a, and exact traction is applied at the top and right side of the mesh, while symmetry boundary conditions are applied at the bottom and left side of the mesh. The analytical solution \cite{Coradello2021} is given as
        \begin{align}
            u_x= & T_x \cdot \frac{R_i}{8 \mu} \left[ \frac{r}{R_i} A
            + 2 \frac{R_i}{r} ( A + B )
            - 2 \frac{R_i^3}{r^3} B\right] \\
            u_y= & T_x \cdot \frac{R_i}{8 \mu} \left[ \frac{r}{R_i} C
            + 2 \frac{R_i}{r} ( D + E ) 
            - 2 \frac{R_i^3}{r^3} E\right], 
        \end{align}
    where
    \begin{equation*}
        \begin{aligned}
            &A= (k+1)\cos(\theta); B= \cos(3\theta); C=(k-3)\sin(\theta)\\
            &D=(1-k)\sin(\theta); E=\sin(3\theta); k=(3-4\nu); \mu=\frac{E}{2(1+\nu)}.
        \end{aligned}
     \end{equation*}
    Figure \ref{fig7}b shows the position of the quadrature points on the cut element, obtained with the tool Algoim \cite{Saye2022}, when the interface is defined by an implicit curve, and by the tool Ginkgo \cite{Antolin2019}, when the interface is defined by a parametric curve. In Figure \ref{fig7}c, the magnitude of the 
    component $u_x$ of the displacement vector $\mathbf{u}=[u_x,u_y]^T$ is shown.
    The graph in Figure \ref{fig9}a shows the comparison of the error with the size of the element, from which it is possible to see that solving the problem with an element trimmed by an implicit curve gives as good results as when solving the problem defining by a parametric interface. In both approaches we see almost the same accuracy.
    
    \subsubsection{Square plate}
    \begin{figure}[t!]
        \centering
        \includegraphics[]{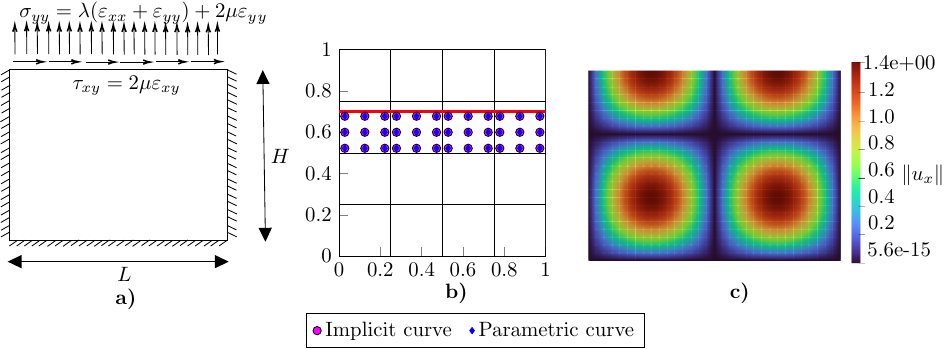}
        \caption{Plane square: a) problem scheme; b) quadrature points; c) displacement magnitude}
        \label{fig8}
    \end{figure}
    \begin{figure}[t!]
        \centering
        \includegraphics[]{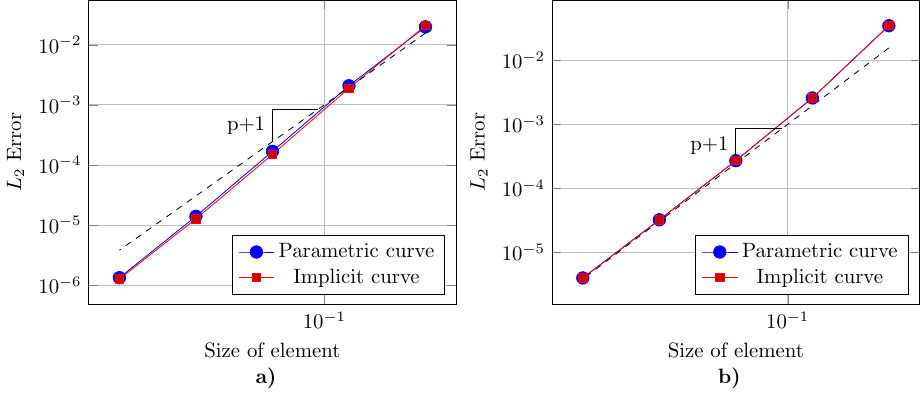}
        \caption{Convergence study for $p$=2: a) infinite plate with a hole; b) square plate}
        \label{fig9}
    \end{figure}
    In this example, we considered a square background mesh $\Omega=[0,1]\times[0,1]$ trimmed at the top, in the position $y=0.75$, by a horizontal straight line, as shown in Figure \ref{fig8}a and \ref{fig8}b, which also shows the quadrature points obtained from both approaches and from the respective open-source codes (Algoim \cite{Saye2022} or Ginkgo \cite{Antolin2019}). The boundary conditions and the source term are manufactured in a such way that the exact solution is given by
       \begin{align}
            u_x=&\sin(2\pi x)\sin(2\pi y)\\
            u_y=&\sin(2\pi x)\sin(2\pi y)
        \end{align}
    Figure \ref{fig8}c shows the magnitude of the component $u_x$ of the displacement field obtained from this problem. Figure \ref{fig9}b presents the graphs of convergence for different element sizes. Both approaches give similar accuracy, which is the same behavior identified on the previous example of an infinite plate with a hole. Hence, we can conclude that both approaches are well-suited for numerical approximation schemes.

\section{CONCLUSIONS}
In this work, we compare two situations of cut elements that occur due to the intersection of embedded interfaces in a background mesh. The two approaches differ in the interface definition, that is, either using parametric or implicit curves. We employ the open-source codes Algoim \cite{Saye2022} and Ginkgo \cite{Antolin2019} to provide the quadrature points and weights over the cut elements. Numerical examples of area computation are defined, from which we can observe similar behaviour in terms of convergence and the number of quadrature points used. Furthermore, using the open-source code GeoPDEs, we apply both approaches to solve 2D elasticity problems with a background mesh defined by B-splines. In this situation, we can compare the behaviour of these procedures not only when the integrand is equal to 1 (area computation problem) but also to observe the effects of the definition of the interface in the solution of a different problem (2D elasticity). In all examples, we observe almost the same accuracy. Therefore, the examples show similar answers for both approaches, which yield that the choice between the type of interface depends mainly on the application of the problem, the advantage of using one rather than the other, and the  interface definition available but not on the numerical aspects of the integration.
\newline
\newline
\textbf{ACKNOWLEDGEMENTS}
\vspace{1mm}
\newline
The work of Guilherme Henrique Teixeira is supported by the joint DFG/FWF Collaborative Research Centre CREATOR (CRC -- TRR361/F90) at TU Darmstadt, TU Graz and JKU Linz/Johann Radon Institute for Computational and Applied Mathematics, Linz.

\bibliographystyle{plainnat}
\bibliography{eccomas2024}

\end{document}